\documentclass[aps,prl,preprint,tightenlines,superscriptaddress,showpacs,byrevtex,subfigure]{revtex4}

\usepackage{graphicx}
\usepackage{color}

\begin{document}
\begin{flushright}
Belle Prerpint 2007-13\\
KEK   Preprint 2006-78
\end{flushright}

\title{ \quad\\[0.5cm] 
Search for Lepton Flavor
Violating $\tau^-$ 
{Decays}\\
into $\ell^-\eta$,  $\ell^-\eta'$ and  $\ell^-\pi^0$ 
}
\affiliation{Budker Institute of Nuclear Physics, Novosibirsk, Russia}
\affiliation{Chiba University, Chiba, Japan}
\affiliation{Chonnam National University, Kwangju, South Korea}
\affiliation{University of Cincinnati, Cincinnati, OH, USA}
\affiliation{The Graduate University for Advanced Studies, Hayama, Japan}
\affiliation{University of Hawaii, Honolulu, HI, USA}
\affiliation{High Energy Accelerator Research Organization (KEK), Tsukuba, Japan}
\affiliation{Hiroshima Institute of Technology, Hiroshima, Japan}
\affiliation{Institute for High Energy Physics, Protvino, Russia}
\affiliation{Institute of High Energy Physics, Vienna, Austria}
\affiliation{Institute for Theoretical and Experimental Physics, Moscow, Russia}
\affiliation{J. Stefan Institute, Ljubljana, Slovenia}
\affiliation{Kanagawa University, Yokohama, Japan}
\affiliation{Korea University, Seoul, South Korea}
\affiliation{Kyungpook National University, Taegu, South Korea}
\affiliation{Swiss Federal Institute of Technology of Lausanne, EPFL, Lausanne, Switzerland}
\affiliation{University of Ljubljana, Ljubljana, Slovenia}
\affiliation{University of Maribor, Maribor, Slovenia}
\affiliation{University of Melbourne, Victoria, Australia}
\affiliation{Nagoya University, Nagoya, Japan}
\affiliation{Nara Women's University, Nara, Japan}
\affiliation{National Central University, Chung-li, Taiwan}
\affiliation{National United University, Miao Li, Taiwan}
\affiliation{Department of Physics, National Taiwan University, Taipei, Taiwan}
\affiliation{H. Niewodniczanski Institute of Nuclear Physics, Krakow, Poland}
\affiliation{Nippon Dental University, Niigata, Japan}
\affiliation{Niigata University, Niigata, Japan}
\affiliation{University of Nova Gorica, Nova Gorica, Slovenia}
\affiliation{Osaka City University, Osaka, Japan}
\affiliation{Osaka University, Osaka, Japan}
\affiliation{Panjab University, Chandigarh, India}
\affiliation{Peking University, Beijing, PR China}
\affiliation{RIKEN BNL Research Center, Brookhaven, NY, USA}
\affiliation{University of Science and Technology of China, Hefei, PR China}
\affiliation{Seoul National University, Seoul, South Korea}
\affiliation{Shinshu University, Nagano, Japan}
\affiliation{Sungkyunkwan University, Suwon, South Korea}
\affiliation{University of Sydney, Sydney, NSW, Australia}
\affiliation{Tata Institute of Fundamental Research, Bombay, India}
\affiliation{Toho University, Funabashi, Japan}
\affiliation{Tohoku Gakuin University, Tagajo, Japan}
\affiliation{Tohoku University, Sendai, Japan}
\affiliation{Department of Physics, University of Tokyo, Tokyo, Japan}
\affiliation{Tokyo Institute of Technology, Tokyo, Japan}
\affiliation{Tokyo Metropolitan University, Tokyo, Japan}
\affiliation{Tokyo University of Agriculture and Technology, Tokyo, Japan}
\affiliation{Virginia Polytechnic Institute and State University, Blacksburg, VA, USA}
\affiliation{Yonsei University, Seoul, South Korea}
\author{Y.~Miyazaki} 
\affiliation{Nagoya University, Nagoya, Japan}
\author{I.~Adachi} 
\affiliation{High Energy Accelerator Research Organization (KEK), Tsukuba, Japan}
\author{H.~Aihara} 
\affiliation{Department of Physics, University of Tokyo, Tokyo, Japan}
\author{D.~Anipko} 
\affiliation{Budker Institute of Nuclear Physics, Novosibirsk, Russia}
\author{K.~Arinstein} 
\affiliation{Budker Institute of Nuclear Physics, Novosibirsk, Russia}
\author{V.~Aulchenko} 
\affiliation{Budker Institute of Nuclear Physics, Novosibirsk, Russia}
\author{T.~Aziz} 
\affiliation{Tata Institute of Fundamental Research, Bombay, India}
\author{A.~M.~Bakich} 
\affiliation{University of Sydney, Sydney, NSW, Australia}
\author{E.~Barberio} 
\affiliation{University of Melbourne, Victoria, Australia}
\author{A.~Bay} 
\affiliation{Swiss Federal Institute of Technology of Lausanne, EPFL, Lausanne, Switzerland}
% ...
\collaboration{The Belle Collaboration}
\noaffiliation

\begin{abstract}
We have searched for {lepton-flavor-violating} $\tau$ 
{decays} 
with a pseudoscalar meson 
{($\eta$, $\eta'$ {and} $\pi^0$)}
using a data sample of 401 fb$^{-1}$ collected 
{with} 
the Belle detector at {the}
KEKB asymmetric-energy $e^+e^-$ collider. 
No evidence for these decays is found {and} 
we set the following upper limits 
{on} the branching fractions:  
${\cal{B}}(\tau^-\rightarrow e^-\eta) < 9.2\times 10^{-8}$, 
${\cal{B}}(\tau^-\rightarrow \mu^-\eta) < 6.5\times 10^{-8}$,
${\cal{B}}(\tau^-\rightarrow e^-\eta') < 1.6\times 10^{-7}$, 
${\cal{B}}(\tau^-\rightarrow \mu^-\eta') < 
1.3\times 10^{-7}${,}  
${\cal{B}}(\tau^-\rightarrow e^-\pi^0) < 8.0\times 10^{-8}$
and 
${\cal{B}}(\tau^-\rightarrow \mu^-\pi^0) < 1.2\times 10^{-7}$    
at the 90\% confidence level.
{These results improve 
{our previously published upper limits
by factors from 2.3 to 6.3.}}
\end{abstract}
\maketitle
 \section{Introduction}

{Lepton flavor violation (LFV)
{appears}
in {various} extensions of the Standard Model (SM),
{e.g.,} 
{supersymmetry (SUSY), leptoquark and many other models.}}
{In particular, $\tau$ {lepton-flavor-violating} decays
with a pseudoscalar 
meson ($M^0 = \eta, \eta'$ and $\pi^0$)
are discussed
in 
models with 
Higgs-mediated LFV processes~\cite{cite:higgs,cite:higgs2,cite:higgs3},
heavy singlet Dirac neutrinos~\cite{cite:amon},
dimension-six effective fermionic operators that induce $\tau-\mu$
mixing~\cite{cite:six_fremionic},
$R-$parity violation in SUSY~\cite{cite:rpv, cite:rpv2, Li:2005rr},
{type III two-Higgs-doublet models~\cite{Li:2005rr} 
{and}
flavor changing $Z'$ bosons~\cite{Li:2005rr}.}
{Some of these models predict branching fractions 
which, for 
certain combinations of model parameters,
can be as high as $10^{-6}$; 
this rate is 
already accessible 
at 
high-statistics 
 $B$ factory experiments.
Previously, we 
obtained 
90\% confidence level (C.L.) upper limits 
for 
various $\tau^-\to\ell^-M^0$ (where $\ell^- = e^-$ or $\mu^-$)
branching fractions 
using 154 fb${}^{-1}$ of data;
the results were 
in the range (1.5$-$10)~$\times~10^{-7}$~\cite{cite:leta}.
The BaBar collaboration
has
recently
used 
339 fb${}^{-1}$ of data
to obtain 90\% C.L. 
upper limits 
in the range (1.1$-$2.4)~$\times~10^{-7}$~\cite{cite:leta_babar}.
Here
we {update}  
our previous results
using  401 fb$^{-1}$ of data.
These {datasets} are 
collected at the $\Upsilon(4S)$ resonance
and 60 MeV below it
with the Belle detector at 
the KEKB  $e^+e^-$
asymmetric-energy collider~\cite{kekb}.}}

The Belle detector is a large-solid-angle magnetic spectrometer that
consists of a silicon vertex detector (SVD),
a 50-layer central drift chamber (CDC),
an array of aerogel threshold Cherenkov counters (ACC), 
a barrel-like arrangement of
time-of-flight scintillation counters (TOF), 
and an electromagnetic calorimeter
{comprised of}
CsI(Tl) {crystals (ECL), all located} inside
a superconducting solenoid coil
that provides a 1.5~T magnetic field.
An iron flux-return located {outside} the coil is instrumented to 
detect $K_{\rm{L}}^0$ mesons
and to identify muons (KLM).
The detector is described in detail elsewhere~\cite{Belle}.

{Particle identification
is very important 
{for this measurement.}
{We use particle identification
likelihood variables based on}}
the ratio of the energy
deposited in the 
ECL to the momentum measured in the SVD and CDC,
the shower shape in the ECL,
the particle range in the KLM,
the hit information from the ACC,
{the $dE/dx$ information in the CDC,}
and {the {particle} time-of-flight} from the TOF.
{{For lepton identification,
we {form} likelihood ratios ${\cal P}(e)$~\cite{EID} 
and ${\cal P}({\mu})$~\cite{MUID}
based on the}
electron and  muon probabilities, respectively,
{which are}
determined by
the responses of the appropriate subdetectors.}

In order to determine the event selection 
{requirements},
we use {Monte Carlo} (MC) samples.
{The following MC programs have been} 
used to
generate background events:
KORALB/TAUOLA~\cite{cite:koralb_tauola} for $\tau^+\tau^-$,
QQ~\cite{cite:qq} for $B\bar{B}$ and continuum,
BHLUMI~\cite{BHLUMI} for {Bhabha events,}
KKMC~\cite{KKMC} for $e^+e^-\rightarrow\mu^+\mu^-$ and
AAFH~\cite{AAFH} for two-photon processes.
Signal MC is generated by KORALB/TAUOLA.
{Signal $\tau$ decays are  two-body
and assumed}
to  have a uniform angular distribution
{in the $\tau$} lepton's rest frame.
{All kinematic variables are
calculated in the laboratory frame
{unless otherwise specified.}
In particular,
variables
calculated in the $e^+e^-$ center-of-mass (CM) frame
are indicated by the superscript ``CM''.}
\section{Event Selection}

{We search for $\tau^+\tau^-$ events
in which one $\tau$ 
decays
{into a lepton and a pseudoscalar meson} 
{on} the signal side,
while 
the other $\tau$ 
decays 
into  one charged track 
with a {sign} opposite to that of {the}
signal-side lepton
and any number of additional photons and neutrinos 
on the tag side.}
Thus, 
the decay chain we reconstruct is:
\begin{center}
{$\left\{
\tau^- \rightarrow \ell^-~(e^-\mbox{ or }\mu^-) + M^0~(\eta,
\eta'\mbox{ or }\pi^0 )
\right\} 
~+
~ \left\{ \tau^+ \rightarrow ({\rm a~track})^+ + (n\geq0~\gamma)
 + X(\rm{missing}) 
\right\}$}\footnotemark[2].
\end{center}
We reconstruct a
pseudoscalar meson
in the following modes:
$\eta\to\gamma\gamma$ and $\pi^+\pi^-\pi^0(\to\gamma\gamma)$,
$\eta'\to\rho(\to\pi^+\pi^-)\gamma$ and $\eta(\to\gamma\gamma)\pi^+\pi^-$,
{and}
$\pi^0\to\gamma\gamma$.
While the $\pi^0\to\gamma\gamma$ and $\eta\to\gamma\gamma$ modes
correspond to {a 1-1 prong} {configuration}, 
the other modes   
{give  
{a 3-1 prong} {configuration}.
{All charged tracks} and photons 
are required to be reconstructed 
{within {the} fiducial {volume}} 
defined by $-0.866 < \cos\theta < 0.956$,
where $\theta$ is the polar angle with
{respect to the direction 
{along}
the $e^+$ beam.}
We select charged tracks with
{momenta} 
{transverse to the $e^+$ beam {direction,}}
$p_t > 0.1$ GeV/$c$
{while 
the photon  energies must satisfy
$E_{\gamma} > 0.1$ GeV} 
($0.05$ GeV) 
{for the 1-1 prong
{(3-1 prong)} configuration.}

\footnotetext[2]{Unless otherwise stated, charge
conjugate decays are
{implied}
throughout
this paper.}

%
%

{Candidate $\tau$-pair events are {thus} 
required to have} 
two and four tracks
{with}  a zero net 
{charge}
for {the} 
1-1 and 3-1 prong 
{configurations,} respectively.
{Event particles {are} separated into two 
hemispheres referred to as {the} signal and 
tag {sides}
using the plane perpendicular to the thrust
axis~\cite{thrust}. 
The tag side
contains a single track,
the signal side contains one or three tracks.}
The track 
{on} the signal side
is required to 
{be identified as a lepton.}
{The electron (muon) {identification} criteria are}
${\cal P}(e)~({\cal P}(\mu)) > 0.9$ with $p >$ 0.7 GeV/c. 
{The efficiencies 
{for} electron and muon 
{identification after} these requirements
are 92\% and 88\%, {respectively.}}
{To {reduce} fake 
pseudoscalar meson {candidates,} 
we reject radiative photons 
from {electrons on} the signal side if $\cos \theta_{e\gamma}>$ 0.99.}

%
%

The $\pi^0$ candidates are formed from 
{pairs} 
of photons
that satisfy
0.115 GeV/$c^2$ $< M_{\gamma\gamma} < $ 0.152 GeV/$c^2${,} 
which corresponds to
$\pm$2.5 standard deviations ($\sigma$)
in terms of the mass resolution.
{We also require} 
{$p_{\pi^0} > 0.1$ GeV/c 
{on} the signal side.}

The $\eta$ meson is
reconstructed 
{in the}
{$\gamma\gamma$ 
{($\pi^+\pi^-\pi^0$)} decay modes.}
The mass window 
is chosen to be
0.515 GeV/$c^2$ (0.532 GeV/$c^2$)  
$< m_{\gamma\gamma} (m_{\pi^+\pi^-\pi^0})< $
0.570 GeV/$c^2$ (0.562 GeV/$c^2$),
{which corresponds to $-$3.0$\sigma$ and $+$2.5$\sigma$ ($\pm3\sigma$ ). 
{To reduce background in $\eta\to\gamma\gamma$ channel,
we reject
those photons
that  form $\pi^0$ candidates
in association with any other {photon} with $E_{\gamma} > 0.05$ GeV,
within
the $\pi^0$ mass window, 
{0.10 GeV/$c^2$ $< M_{\gamma\gamma} < 0.16$ GeV/$c^2$}.}

The $\eta'$ meson is
reconstructed 
{in the}
{$\rho\gamma$ and $\eta\pi^+\pi^-$ decay modes.}
For the  $\rho\to\pi^+\pi^-$ {selection,}
the mass window is chosen to be
0.550 GeV/$c^2$ $< m_{\pi\pi} < 0.900$ GeV/$c^2$.
{We} reconstruct {$\eta'$ candidates} 
using a $\rho$ candidate and {a} photon
on the signal side.
The $\eta'$ mass window is chosen to be
$0.930$ GeV/$c^2$ $< m_{\rho\gamma} < 0.970$ GeV/$c^2$,
{which corresponds to {$-3.0\sigma$} and $+$2.5$\sigma$.}
Furthermore,
{we veto photons from $\pi^0$ candidates}
in order to avoid fake $\eta'$ candidates
from  $\pi^0\to\gamma\gamma$.
We remove events if a $\pi^0$ {candidate}
with
{invariant mass in the range}
0.10 GeV/$c^2$ $< M_{\gamma\gamma} < 0.16$ GeV/$c^2$
is reconstructed 
{using} a photon 
from the $\eta'$ candidate 
and another photon  with $E_{\gamma} > 0.05$ GeV.
Figure~\ref{fig:muetap_rhogam}
shows the $\rho\to\pi^+\pi^-$  and
$\eta'\to\rho\gamma$ mass distributions.
{The} dominant backgrounds
for this mode 
come from $\tau^-\to h^-\rho^0\nu_{\tau}(+\pi^0)$
{(where $h^- = K^- \mbox{ or } \pi^-$)}
{with} 
{a photon} from
{$\pi^0$} decay, beam background 
{or} {initial state radiation (ISR).}
{{As shown in Fig.~\ref{fig:muetap_rhogam},
{there is} no $\eta'$ peak 
either {in}  data or in  MC}
since 
{decay modes with {an} 
$\eta'$ are very rare}
and {are} not included 
{in the} generic $\tau$ decay {model}~\cite{PDG}.}
{We 
{also}
reconstruct  $\eta'$ {candidates} 
using an $\eta$ candidate and two oppositely charged tracks 
consistent with being {pions.}}
{We 
{impose a}
$\gamma\to e^+e^-$ 
{conversion veto} 
as $P(e) < 0.1$ \text{for both tracks} 
{in} the $\eta'$ candidate.}
The $\eta$ products from $\eta'\to\eta\pi^+\pi^-$ decay
{are}
reconstructed 
{using} two photons.
The $\eta'$ mass window is chosen to be
$0.920$ GeV/$c^2$ $< m_{\eta\pi^+\pi^-} < 0.980$ GeV/$c^2$,
{which corresponds to $\pm$3.0$\sigma$.}

\begin{figure}
 \resizebox{0.8\textwidth}{0.4\textwidth}{\includegraphics
 {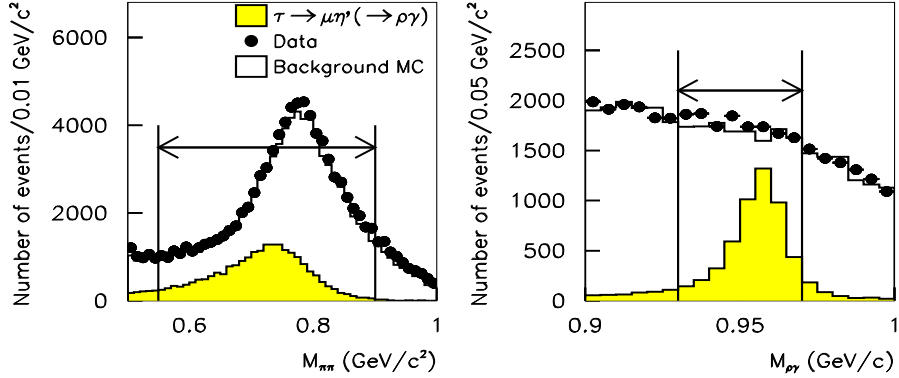}}
 \vspace*{-0.5cm}
\caption{{The $\rho\to\pi^+\pi^-$ (left) and
$\eta'\to\rho\gamma$ (right) mass
 {distributions.}
 While the signal MC ($\tau^-\rightarrow\mu^-\eta'$)
 distribution is normalized arbitrarily,
 {the data and background MC} are normalized to the same luminosity.
 Selected regions are indicated  
 by the
 arrows from the marked cut boundaries.} }
\label{fig:muetap_rhogam}
\end{figure}

To ensure that the missing particles are neutrinos rather
than photons or charged particles
{that fall outside the detector acceptance,}
we impose additional requirements on the missing
momentum vector 
$\vec{p}_{\rm miss}$,
{which is} calculated by subtracting the
vector sum of the momenta
of
all tracks and photons
from the sum of the $e^+$ and $e^-$ beam momenta.
We require that the magnitude of $\vec{p}_{\rm miss}$
{be} greater than
{0.4 GeV/$c$,}
and that
{its direction 
{point} into}
the fiducial volume of the
detector.
{Since neutrinos are 
{normally}
emitted only on the tag side,
the direction of
{$\vec{p}_{\rm miss}$
should lie within the tag side of the event.}}
The cosine of the
opening angle between
{$\vec{p}_{\rm miss}$}
and
{the} thrust axis ({on} the signal side)
in the CM system,
{$\cos \theta^{\mbox{\rm \tiny CM}}_{\rm miss-thrust}$,}
{is therefore required to be less than {$-0.55$}}.

%
%

For {the 1-1 prong} configuration,
to suppress fake $\eta$ ($\pi^0$) 
{candidates arising from}
{beam background and {ISR,}}  
we require that the higher and lower energy  {photons} 
($E_{\gamma1}$ and $E_{\gamma2}$)
in an $\eta$ 
{($\pi^0$)} 
candidate 
{satisfy the requirement} 
$E_{\gamma1} > $ 0.6 (0.9) GeV and 
$E_{\gamma2} >$ 0.25 (0.2) GeV, respectively,
as shown for {the} $\tau^-\rightarrow\mu^-\pi^0$ mode
in 
{Figs.~\ref{fig:cut_pi0}} (a) and (b).
{In order to suppress background
from $q\bar{q}$ ($q = u, d, s, c$)
continuum events,
{we require that 
{the} number 
of extra photon candidates on the signal and tag side
($n_{\gamma}^{\rm{SIG}}$ and $n_{\gamma}^{\rm{TAG}}$)
be $n_{\gamma}^{\rm{SIG}}\leq1$
and
$n_{\gamma}^{\rm{TAG}}\leq2$, respectively.}
{To reduce background} 
from Bhabha and $\mu^+\mu^-$ events,
we require the momentum of 
{the} 
lepton and 
{that of}
{the}
tag-side charged particle in the CM system
{to be} less than 4.5 {GeV/$c$}.
Furthermore,
we require the momentum of a lepton to be
greater than 1.5 GeV/$c$
for {the}
$\tau^-\to\ell^-\pi^0$ mode
({listed} in Fig.~\ref{fig:cut_pi0} (c)).
{This condition is not 
imposed for the $\tau^-\to\ell^-\eta$ mode,
in which {the average lepton momentum} 
as well
as the background level are 
{lower.}}

%
%
The reconstructed mass {on} 
the tag side using a track 
(with {a} pion mass hypothesis)
and {photons},
$m_{\rm tag}$, 
{is required to be} 
less than 1.777 GeV/$c^2$.
The total visible energy in the CM frame{,}
{$E^{\mbox{\rm{\tiny{CM}}}}_{\rm{vis}}$,}
is defined as the sum of the energies
{of the $\eta$ candidate,
the lepton,
the tag-side track
(with {a} pion mass hypothesis)
and all photon candidates.
{We require $E^{\mbox{\rm{\tiny{CM}}}}_{\rm{vis}}$
to satisfy}
{the condition
shown in Table~\ref{tbl:eachcut}.
To reduce background from $\mu^+\mu^-$, 
{two-photon} and Bhabha events,
{we 
{add}
{the veto condition}
$E^{\mbox{\rm{\tiny{CM}}}}_{\rm{vis}} > 8.5$ GeV 
for {the muon (electron) mode} 
if the track {on} the tag side is {a} muon (electron).}
The cosine of the
opening angle between 
the lepton
{and the} $M^0$
in the CM system,
{$\cos \theta^{\mbox{\rm \tiny CM}}_{\rm \ell-M^0}$,}
{is required to 
{lie in the range} 
shown in Table~\ref{tbl:eachcut}
({also} shown 
in Fig.~\ref{fig:cut_pi0} (d) 
{for 
{the}
$\tau^-\to\mu^-\pi^0$ mode).}
{For all kinematic distributions {in Fig.~\ref{fig:cut_pi0},}
{reasonable agreement between the data and background MC is observed.}}

\begin{figure}
\begin{center}
       \resizebox{0.8\textwidth}{0.8\textwidth}{\includegraphics
        {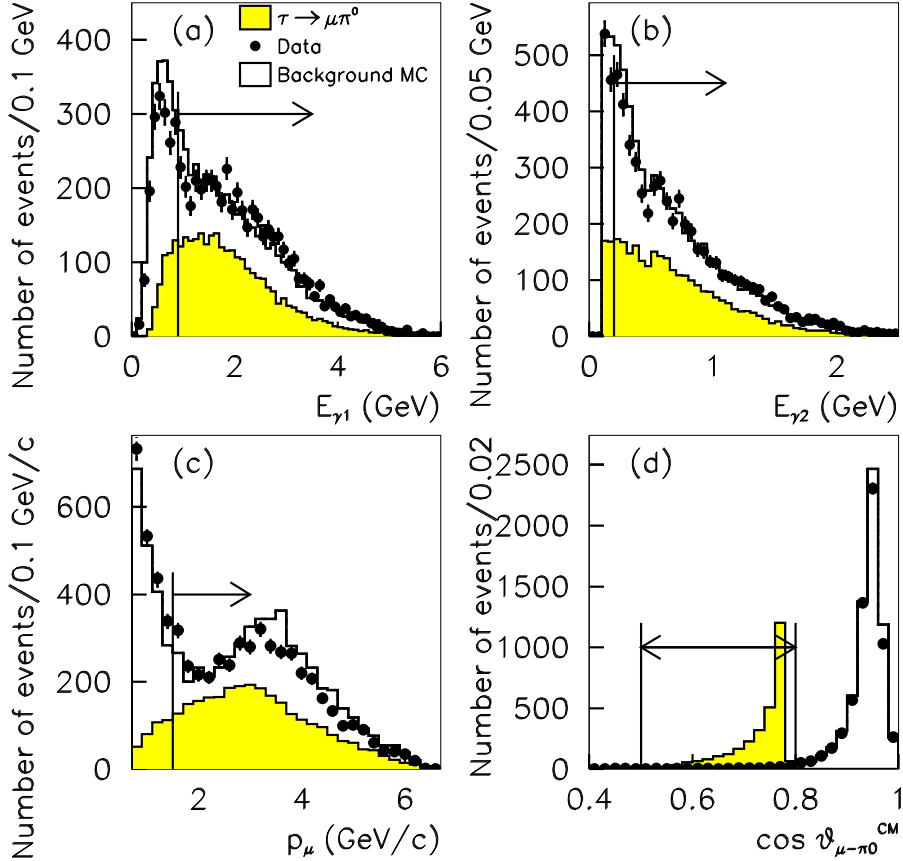}}
 \vspace*{-0.5cm}
 \caption{ 
 {Kinematic distributions used in the event selection:
 (a) higher energy and
 (b) lower energy of a  photon from the $\pi^0$ candidate 
  ($E_{\gamma1}$ and $E_{\gamma2}$);
 (c) momentum of a muon ($p_{\mu}$);
 (d) the cosine of the opening angle between the muon and
 $\pi^0$ in the CM frame ($\cos \theta_{\mu-\pi^0}^{\rm CM}$).
 While the signal MC {($\tau^-\to\mu^-\pi^0$)}
 distribution is normalized arbitrarily, 
 {the data and background MC} are normalized to the same luminosity.
 {Selected regions are indicated  
 by {the}
 arrows from the marked cut {boundaries.}}} 
}
\label{fig:cut_pi0}
\end{center}
\end{figure}

%
%

The correlation between {the}
momentum of the track on the tag side,
$p^{\rm CM}_{{\rm tag}}$,
and
{the} cosine of the opening angle
{between the thrust and missing particle,}
$\cos \theta^{\rm CM}_{{\rm miss-thrust}}$
{in the CM system}
{is employed to further suppress
{backgrounds} 
from generic $\tau^+\tau^-$
and $\mu^+\mu^-$ events via the following requirements:}
$p^{\rm CM}_{{\rm tag}} > 1.1\log(\cos \theta^{\rm CM}_{{\rm miss-thrust}}+0.92)+5.5$,
and 
$p^{\rm CM}_{{\rm tag}} < 5\cos \theta^{\rm CM}_{{\rm miss-thrust}}+7.8$
where $p^{\rm CM}_{\rm tag}$ is in GeV/$c$
(see Fig.~\ref{fig:tag-cos}). 
Finally, 
we require
the following relation
between the missing momentum $p_{\rm{miss}}$ and
missing mass squared
$m^2_{\rm{miss}}$ 
{to} 
further suppress background from generic $\tau^+\tau^-$
and
continuum background.
In signal events, 
two neutrinos are included 
if the $\tau$ decay {on} the tag side is 
{a} leptonic decay,
while 
one neutrino is included if the $\tau$ decay {on} the tag
side is {a} hadronic decay.
{{Therefore,} 
we 
separate events into
{two classes}
according to the track {on} the tag side: leptonic or hadronic,}
{{and apply} the requirements 
{shown in Table~\ref{tbl:misscut}}
{(see also Fig.~\ref{fig:pmiss_vs_mmiss2}). }}

\begin{figure}
\begin{center}
\resizebox{0.8\textwidth}{0.8\textwidth}{\includegraphics 
{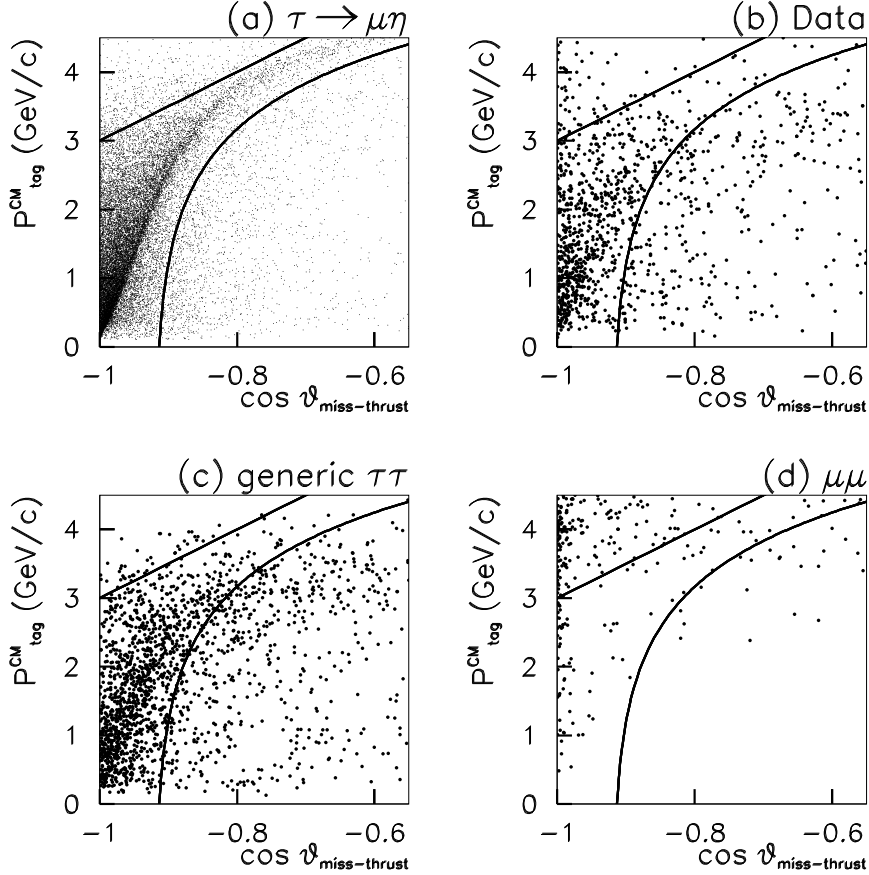}}
 \vspace*{-1cm}
 \caption{
{Scatter-plots 
for  (a) signal MC ($\tau^-\rightarrow\mu^-\eta(\to\gamma\gamma)$),  
(b) data,  
(c) generic $\tau^+\tau^-$ MC events 
 and (d) $\mu^+\mu^-$ MC events in {the}
$p^{\rm CM}_{\rm tag}-\cos \theta_{\rm miss-thrust}$ plane.
 {Selected regions lie between {the two curves}.}}
 \label{fig:tag-cos}
 }
 \end{center}
\end{figure}

\begin{figure}
\begin{center}
 \resizebox{0.8\textwidth}{0.8\textwidth}{\includegraphics
 {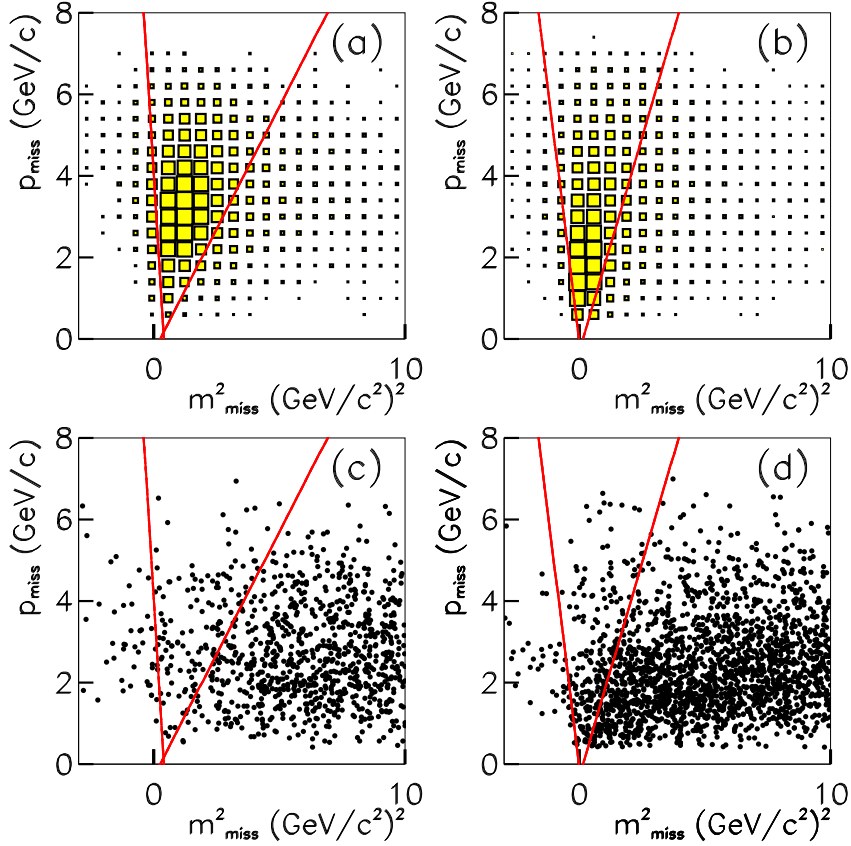}}
 \vspace*{-0.5cm}
 \caption{
Scatter-plots of 
$p_{\rm miss} -   
m_{\rm miss}^2$,
for leptonic 
and  hadronic tags:
{ (a) shows
the signal MC ($\tau^-\to\mu^-\eta$) 
distribution
with arbitrary normalization
for a leptonic tag
while (c) shows
the data distribution for a leptonic tag;
(b) shows
the signal MC ($\tau^-\to\mu^-\eta$) 
distribution
with arbitrary normalization
for a hadronic  tag
while Fig. (d) shows
the data distribution for a hadronic tag.}
Selected regions are indicated by lines.}
\label{fig:pmiss_vs_mmiss2}
\end{center}
\end{figure}

\begin{table}
\begin{center}
\begin{tabular}{|c|c|c|} \hline
 & 
 $E^{\mbox{\rm{\tiny{CM}}}}_{\rm{vis}}$  & 
 $\cos \theta^{\mbox{\rm \tiny CM}}_{\rm \ell-M^0}$ \\ \hline
$\tau^-\to\ell^-\eta$ &  
0.529 GeV $< E^{\mbox{\rm{\tiny{CM}}}}_{\rm{vis}} < $10.0 GeV &
$0.50 < \cos \theta^{\mbox{\rm \tiny CM}}_{\rm \ell-\eta} < 0.85$ \\
$\tau^-\to\ell^-\eta'$ &  
0.529 GeV $< E^{\mbox{\rm{\tiny{CM}}}}_{\rm{vis}} < $11.0 GeV &
$0.50 < \cos \theta^{\mbox{\rm \tiny CM}}_{\rm \ell-\eta'}$ \\
$\tau^-\to\ell^-\pi^0$ &  
0.529 GeV $< E^{\mbox{\rm{\tiny{CM}}}}_{\rm{vis}} < $10.0 GeV &
$0.50 < \cos \theta^{\mbox{\rm \tiny CM}}_{\rm \ell-\pi^0} < 0.80$ \\
\hline
\end{tabular}
\caption{
The selection criteria 
{for} 
the total visible energy in the CM frame
($E^{\mbox{\rm{\tiny{CM}}}}_{\rm{vis}}$) 
and
the cosine of the
opening angle between 
the lepton
{and the} 
$M^0$
in the CM system
($\cos \theta^{\mbox{\rm \tiny CM}}_{\rm \ell- M^0}$).
}
\label{tbl:eachcut}
\end{center}
\end{table}

\begin{table}
\begin{center}
\begin{tabular}{|c|c|c|} \hline
 & {Leptonic mode}  &{Hadronic mode} 
\\ \hline
$\eta\to\gamma\gamma$ & 
\begin{tabular}{c}
  $p_{\rm miss} > -10m^2_{\rm miss}+4$  \\
  $p_{\rm miss} > 1.1m^2_{\rm miss}-0.3$ \\
\end{tabular}
&
\begin{tabular}{c}
  $p_{\rm miss} > -5m^2_{\rm miss}-0.25$ \\ 
  $p_{\rm miss} > 2.1m^2_{\rm miss}-0.3$ \\ 
\end{tabular} \\\hline
$\eta\to\pi^+\pi^-\pi^0$ & 
\begin{tabular}{c}
$p_{\rm miss} > -10m^2_{\rm miss}-4$\\
$p_{\rm miss} > 1.1m^2_{\rm miss}-1$\\ 
\end{tabular}
&
\begin{tabular}{c}
 $p_{\rm miss} > -5m^2_{\rm miss}-0.25$ \\ 
 $p_{\rm miss} > 2.1m^2_{\rm miss}-0.3$ \\
\end{tabular}\\\hline
$\eta'\to\rho\gamma$ & 
\begin{tabular}{c}
$p_{\rm miss} > -8m^2_{\rm miss}-0.2$ \\
$p_{\rm miss} > 1.2m^2_{\rm miss}-0.3$ \\
\end{tabular}
&
\begin{tabular}{c}
$p_{\rm miss} > -5m^2_{\rm miss}-0.2$ \\
$p_{\rm miss} > 2m^2_{\rm miss}-0.3$  \\
\end{tabular}\\\hline
$\eta'\to\eta\pi^+\pi^-$ & 
\begin{tabular}{c}
$p_{\rm miss} > -3m^2_{\rm miss}$ \\
$p_{\rm miss} > 1.5m^2_{\rm miss}-0.5$ \\ 
\end{tabular}
&
\begin{tabular}{c}
$p_{\rm miss} > -4m^2_{\rm miss}-0.8$ \\
$p_{\rm miss} > 2.5m^2_{\rm miss}-0.2$ \\
\end{tabular}\\\hline
$\pi^0\to\gamma\gamma$ & 
\begin{tabular}{c}
  $p_{\rm miss} > -10m^2_{\rm miss}+4$  \\
  $p_{\rm miss} > 1.1m^2_{\rm miss}-0.3$ \\
\end{tabular}
&
\begin{tabular}{c}
  $p_{\rm miss} > -5m^2_{\rm miss}-0.25$ \\ 
  $p_{\rm miss} > 2.1m^2_{\rm miss}-0.3$ \\ 
\end{tabular} \\\hline
\end{tabular}
\caption{
The selection criteria {for} 
the missing momentum ($p_{\rm{miss}}$) and
missing mass squared ($m^2_{\rm miss}$) 
{correlations,}
{$p_{{\rm miss}}$} is in GeV/$c$
and  $m^2_{\rm miss}$ is in $(\rm{GeV}/c^2)^2$.
}
\label{tbl:misscut}
\end{center}
\end{table}

For {the}
3-1 prong configuration,
we
{impose  
similar 
requirements}
$m_{\rm {tag}} < $ 1.777 GeV/$c^2$,
{$n_{\gamma}^{\rm{SIG}}\leq 1$,}
$E^{\mbox{\rm{\tiny{CM}}}}_{\rm{vis}}$ 
and
$\cos \theta^{\mbox{\rm \tiny CM}}_{\rm \ell-M^0}$ 
{(Table~\ref{tbl:eachcut})}.
For {the}
$\eta'\to\rho\gamma$ mode,
we  {require} the photon energy to
be greater than
0.25 GeV for the barrel 
and  
0.40 GeV for the forward region
to suppress 
{fake candidates} 
from 
beam background
and
ISR.
{The 
cut 
{on}
the
{$n_{\gamma}^{\rm{TAG}}$} and 
$p^{\rm CM}_{{\rm tag}}$-$\cos \theta^{\rm CM}_{{\rm miss-thrust}}$
correlation is not 
applied for the 3-1 configuration modes,
in which the background level is 
{lower}
than 
{in} 
the 1-1 configuration modes.}
{We apply 
{requirements} 
on $p_{\rm miss}$ and $m^2_{\rm miss}$ similar to 
those for the 1-1 prong configuration
{(see Table~\ref{tbl:misscut}).}}

\section{Signal Region and Background Estimation}

Signal candidates are examined in the two-dimensional
{plots}
of the $\ell^-M^0$ invariant
mass, $M_{\rm {inv}}$, and 
the difference of their energy from the
beam energy in the CM system, $\Delta E$.
A signal event should have $M_{\rm {inv}}$
close to the $\tau$-lepton mass
and
$\Delta E$ close to {zero.}
{For all modes,
the $M_{\rm {inv}}$ and $\Delta E$  resolutions} are parameterized
from the MC distributions 
{{with  asymmetric Gaussian shapes}
to account for initial state radiation and ECL energy leakage for photons.
The resolutions in
$M_{\rm inv}$ and $\Delta E$ are 
{given}
in Table~\ref{tbl:reso_del_e_m}.

\begin{table}
\begin{center}
\begin{tabular}{c|cc|cc} \hline 
Mode
& $\sigma^{\rm{high}}_{M_{\rm{inv}}}$ (MeV/$c^2$)  
& $\sigma^{\rm{low}}_{M_{\rm{inv}}}$ (MeV/$c^2$)
& $\sigma^{\rm{high}}_{\Delta E}$ (MeV)      
&  $\sigma^{\rm{low}}_{\Delta E}$ (MeV)
 \\ \hline
$\mu\eta(\to\gamma\gamma)$ 
& 14.7 &  19.4 & 30.3  & 61.4  \\  
$\mu\eta(\to\pi^+\pi^-\pi^0)$ 
& 7.2  & 8.5 & 18.5 & 36.4 \\  \hline
$e\eta(\to\gamma\gamma)$ 
& 14.0  &  19.8 & 37.3 & 62.4  \\  
$e\eta(\to\pi^+\pi^-\pi^0)$ 
 & 7.6 &  9.3 & 19.4 & 41.8 \\  \hline\hline
$\mu\eta'(\to\rho\gamma)$
 & 7.8  &  9.0 & 16.8  & 34.1 \\

$\mu\eta'(\to\eta\pi^+\pi^-)$
& 11.2  & 19.1 & 27.1 & 53.5  \\  \hline

$e\eta'(\to\rho\gamma)$
&  9.2  &  10.4 & 19.6 & 40.0  \\

$e\eta'(\to\eta\pi^+\pi^-)$
 & 10.3  &  21.9 & 26.1 & 59.4  \\  \hline\hline

$\mu\pi^0(\to\gamma\gamma)$
& 14.9  &  19.1 & 33.8 & {63.0} \\
$e\pi^0(\to\gamma\gamma)$
 & 12.7  & 23.1 & 35.6 & 64.6  \\  \hline

\end{tabular}
\caption{Summary {of} $M_{\rm inv}$ (MeV/$c^2$) and 
$\Delta E$ 
{resolutions} (MeV)}
\label{tbl:reso_del_e_m}
\end{center}
\end{table}

{To evaluate the branching {fractions,} 
we use  {elliptical signal regions,} 
which {contain} 90\% 
of the MC signal {events satisfying} all cuts.
{These} 
signal regions are shown in Fig.~\ref{fig:openbox} and 
{\ref{fig:openbox2}};
the corresponding signal {efficiencies} are given in Table~\ref{tbl:eff}.}
We blind {the signal region}
{so as not to bias our choice of selection criteria.}
Figures~\ref{fig:openbox} 
and \ref{fig:openbox2} show 
scatter-plots
for data and signal MC samples 
distributed over $\pm 10\sigma$
in the $M_{\rm{inv}}-\Delta E$ plane.
{Most of the surviving
background events in 
$\tau\to\ell\pi^0$ modes 
{come from 
$\tau^-\to\pi^-\pi^0\nu_{\tau}$,}
{where} 
{{the} 
$\pi^-$ 
{is misidentified as a lepton.}}
The remaining backgrounds 
in the $\tau^-\rightarrow\mu^-\eta(\to\gamma\gamma)$ mode
are from
{$\tau$ decay including 
{a} real $\eta$ meson
or combinations of a fake lepton and a fake $\eta$ meson
formed by $\gamma$'s from $\pi^0$ decay,
ISR 
{or} 
beam background.}
{As there are few remaining MC background events 
in the signal ellipse, 
we estimate the background contribution 
using the $M_{\rm{inv}}$  sideband regions.
Extrapolation to the signal region assumes that
the background distribution  is
flat along the $M_{\rm inv}$ axis.}
We then estimate {the} expected number of the background events in the signal
region for each  mode
using  the number of 
{data events observed} 
in the sideband region  
inside 
{the} horizontal lines {but}
excluding the signal region
as shown in Fig.~\ref{fig:openbox} and \ref{fig:openbox2}.
The numbers of background events in the 90\% elliptical signal region 
{are} {also}
shown in Table~\ref{tbl:eff}.

\begin{table}
\begin{center}
\begin{tabular}{|c|c|c|c|c|c|c|}\hline 
Mode & ${\cal{B}}_{M^0}$ & $\varepsilon$ {(\%)} & 
 $b_0$  &  $s$ & Total Sys. {(\%)} & $s_{90}$ \\ \hline
$\tau\to\mu\eta(\to\gamma\gamma)$ & 0.3938 & 6.42 & 0.40$\pm$0.29 & 0 &
 7.1 & 2.1 \\
$\tau\to\mu\eta(\to\pi^+\pi^-\pi^0)$ & 0.227 & 6.84 & 
 0.24$\pm$0.24 & 0 & 5.6 & 2.2  \\ \hline
$\tau\to e\eta(\to\gamma\gamma)$ & 0.3938 & 4.57 & 0.25$\pm$0.25 
 & 0 & 7.1 & 2.2 \\
$\tau\to e\eta(\to\pi^+\pi^-\pi^0)$ & 0.227 & 4.72 & 0.53$\pm$0.53
 & 0 & 5.6 & 2.0   \\ \hline\hline
$\tau\to\mu\eta'(\to\rho\gamma)$ & 0.294$\times$1.0 & 5.40 &
 0.23$\pm$0.23 & 0 & 6.8 &  2.2 \\
$\tau\to\mu\eta'(\to\eta\pi^+\pi^-)$ & 0.445$\times$0.3943 & 4.92 &
 0.0${}^{+0.23}_{-0.0}$ & 0 & 8.3 & 2.5  \\ \hline
$\tau\to e\eta'(\to\rho\gamma)$ & 0.294$\times$1.0 & 4.76 & 0.0${}^{+0.33}_{-0.0}$
 & 0 & 6.8 & 2.5 \\
$\tau\to e\eta'(\to\eta\pi^+\pi^-)$ & 0.445$\times$0.3943 & 4.27 & 0.0${}^{+0.24}_{-0.0}$
 & 0 & 8.3 & 2.5  \\ \hline\hline
$\tau\to\mu\pi^0(\to\gamma\gamma)$ & 0.98798 & 4.53 & 0.58$\pm$0.34 & 1
 & 4.5 & 3.8\\
$\tau\to e\pi^0(\to\gamma\gamma)$ & 0.98798 & 3.93 & 0.20$\pm$0.20
 & 0 & 4.5 & 2.2 \\
\hline
\end{tabular}
\caption{Results of the final event selection for 
the individual modes:
${\cal{B}}_{M^0}$ is the branching fraction for the $M^0$ decay;
$b_0$ and $s$ are the number of expected background and 
observed events in the signal region, respectively;
``Total sys.'' means the total systematic uncertainty; 
{$s_{90}$} is the upper limit on the number of signal events
including systematic uncertainties.
}
\label{tbl:eff}
\end{center}
\end{table}

Systematic uncertainties for  $M^0$ reconstruction 
{are estimated to be}
3.0\%, 4.0\%, 4.0\%, 5.0\% and 3.0\%
for 
$\eta\to\gamma\gamma$, $\eta\to\pi^+\pi^-\pi^0$,
$\eta'\to\rho\gamma$, 
{$\eta'\to\eta\pi^+\pi^-$} and 
$\pi^0\to\gamma\gamma$,
respectively. 
Furthermore,
the uncertainties due to the branching 
{fractions} of the $M^0$ meson are
{0.7\%,} 1.8\%, {3.1\% and 3.1\%} 
for $\eta\to\gamma\gamma$,  $\eta\to\pi^+\pi^-\pi^0$,
$\eta'\to\rho\gamma$ and  $\eta'\to\eta\pi^+\pi^-$, respectively
~\cite{PDG}. 
For the $\pi^0$ veto  we 
{assign} a 5.5\% uncertainty  
{for} {the $\eta\to\gamma\gamma$ mode}
while a 2.8\% uncertainty  
is assigned {to} 
the $\eta'\to\rho\gamma$ mode. 
The uncertainties {in the} trigger (0.5$-$1.0\%), 
{tracking for lepton on the signal side and 
track on the tag side (1.0\% per each track),} 
{lepton identification (2.0\%),}
MC statistics (1.0$-$1.5\%) 
{and} 
luminosity (1.4\%) are also considered. 
{All these uncertainties are added in quadrature, 
and the total systematic uncertainties are
shown in Table~\ref{tbl:eff}.}  

While the angular distribution of signal $\tau$ decays is 
initially assumed to be uniform {in this analysis},
it is sensitive to the {lepton-flavor-violating} interaction
structure~\cite{LFV}.
The spin correlation 
between the $\tau$ lepton {on} the signal and that {on}  the tag side
must be considered.
A possible nonuniformity {is} 
taken into account by comparing
the uniform case with {MC's}
assuming $V-A$ and $V+A$ interactions,
which result in the maximum possible variations.
No statistically significant difference in 
the $M_{\rm inv}$ -- $\Delta{E}$
distribution or the efficiencies is found {compared to}
the case of the uniform distribution.
Therefore,
systematic uncertainties due to these effects 
are neglected in {the} upper limit evaluation.

{We open the blind and find 
{only 
one event  in 
$\tau\to\mu\pi^0(\to\gamma\gamma)$.
In other modes,
no events} 
are found
in the {blinded} region.}
Since no statistically significant excess of data over
the expected background in the signal region is observed,} 
we set upper limits for branching fractions.
The upper limit {on the number of} signal events 
at the 90\% C.L. $s_{90}$ 
including  systematic uncertainty 
is obtained 
with the use of the Feldman-Cousins method~\cite{cite:FC}
calculated 
by the POLE program without conditioning \cite{pole}.   
The upper limit {on} the branching fraction ($\cal{B}$) is 
{then given by}
\begin{equation}
{{\cal{B}}(\tau^-\to\ell^- M^0) <
\displaystyle{\frac{s_{90}}{2N_{\tau\tau}\varepsilon{{\cal{B}}_{M^0}}}},}
\end{equation}
where
{${\cal B}_{M^0}$ {is taken} from 
{Ref.}~\cite{PDG}}
and
{$N_{\tau\tau} =  357.7\times 10^6$
is
{the number of $\tau^+\tau^-$pairs
{produced} in
401 fb${}^{-1}$ of data.
We obtain $N_{\tau\tau}$
using
$\sigma_{\tau\tau} = 0.892 \pm 0.002$ nb,
the $e^+e^- \rightarrow \tau^+\tau^-$ cross section
at the $\Upsilon(4S)$ resonance
calculated by KKMC~\cite{KKMC}.}}
{The combined upper limits 
{for the}
$\eta$ and $\eta'$ modes
are 
{obtained}
by summing 
$\epsilon\cal{B}$, 
the 
observed 
and expected background events 
of each subdecay,
and systematic uncertainties are 
{estimated} 
by 
first summing all correlated terms linearly and 
then 
{adding quadratically}  
the uncorrelated terms.}
{The upper limits for the branching fractions 
${\cal{B}}(\tau^-\rightarrow \ell^- M^0)$ 
are
{in the range}
$(6.5-16)\times 10^{-8}$} 
at the 90\% confidence level. 
{A} summary {of} the upper limits is 
{given} in Table~\ref{tbl:results}.
{{These results improve  
our previously 
published upper limits}~\cite{cite:leta} 
by factors of 2.3$-$6.3.
{They also improve
upon the recent BaBar results~\cite{cite:leta_babar}  
by
factors of $\sim$ 
1.5.}

\begin{table}
\begin{center}
\begin{tabular}{|l|l|c|}\hline
Mode & $M^0$ subdecay mode & Upper limit 
{on} $\cal B$ at 90\% C.L. \\\hline
$\tau^-\to \mu^-\eta$ & $\eta\to\gamma\gamma$ & 1.2$\times 10^{-7}$ \\
                  & $\eta\to\pi^+\pi^-\pi^0$ & 2.0$\times 10^{-7}$ \\
                  & Combined  & 6.5$\times 10^{-8}$ \\\hline 
$\tau^-\to e^-\eta$ & $\eta\to\gamma\gamma$ & 1.7$\times 10^{-7}$ \\
                  & $\eta\to\pi^+\pi^-\pi^0$ & 2.6$\times 10^{-7}$ \\
                  & Combined  & 9.2$\times 10^{-8}$ \\\hline \hline 
$\tau^-\to \mu^-\eta'$ & $\eta'\to\rho\gamma$ & 1.9$\times 10^{-7}$ \\
                  & $\eta'\to\eta\pi^+\pi^-$ & 4.1$\times 10^{-7}$ \\
                  & Combined  & 1.3$\times 10^{-7}$ \\\hline 
$\tau^-\to e^-\eta'$ & $\eta'\to\rho\gamma$ & 2.5$\times 10^{-7}$ \\
                  & $\eta'\to\eta\pi^+\pi^-$ & 4.7$\times 10^{-7}$ \\
                  & Combined  & 1.6$\times 10^{-7}$ \\\hline \hline 
$\tau^-\to \mu^-\pi^0$ & $\pi^0\to\gamma\gamma$ & 1.2$\times 10^{-7}$ \\
\hline 
$\tau^-\to e^-\pi^0$ & $\pi^0\to\gamma\gamma$ & 8.0$\times 10^{-8}$ \\
\hline 
\end{tabular}
\caption{Summary {of} upper limits {on} $\cal B$ at {90\% C.L.}}
\label{tbl:results}
\end{center}
\end{table}

\begin{figure}
\begin{center}
 \resizebox{0.3\textwidth}{0.3\textwidth}{\includegraphics
 {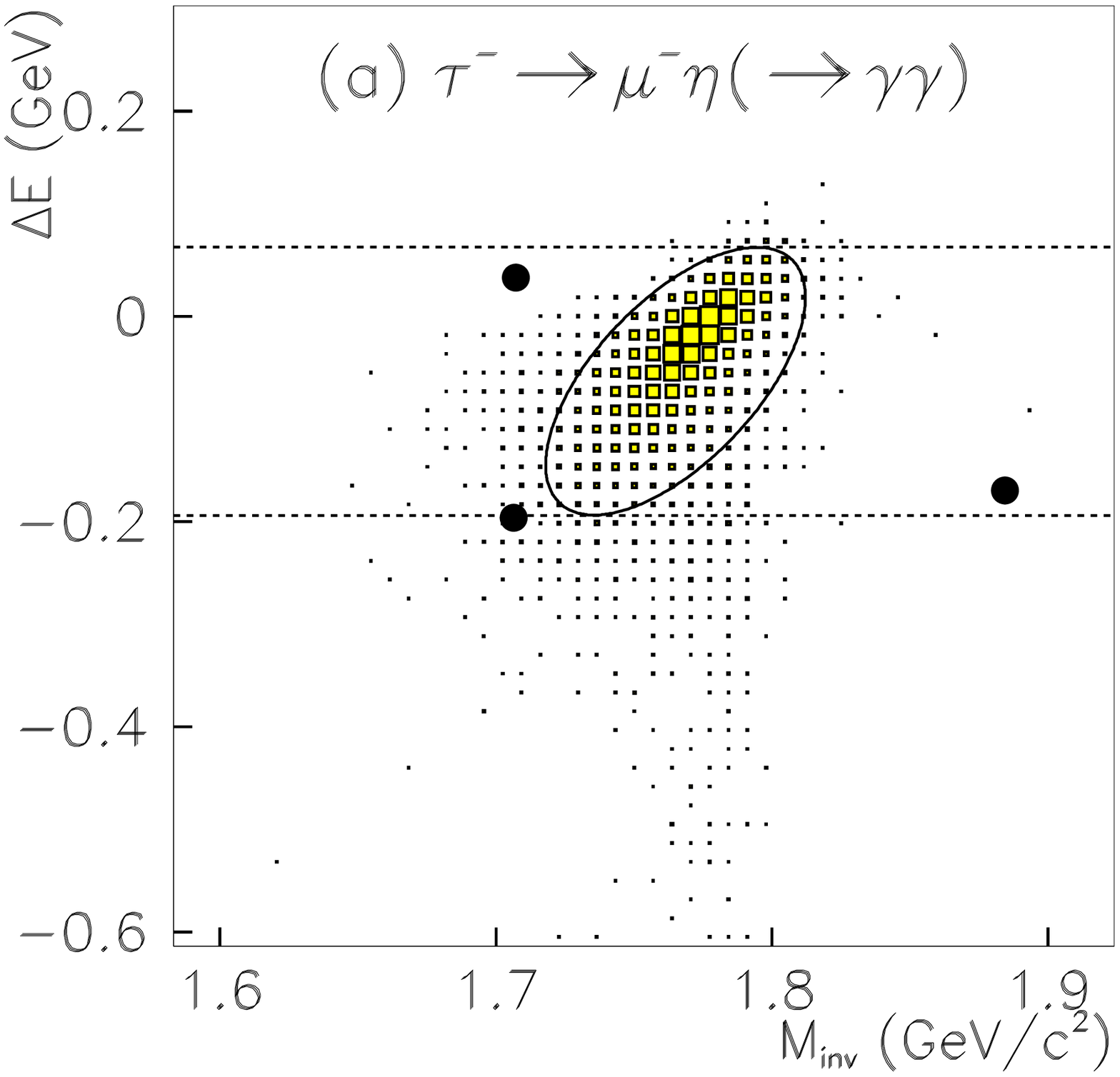}}
 \resizebox{0.3\textwidth}{0.3\textwidth}{\includegraphics
 {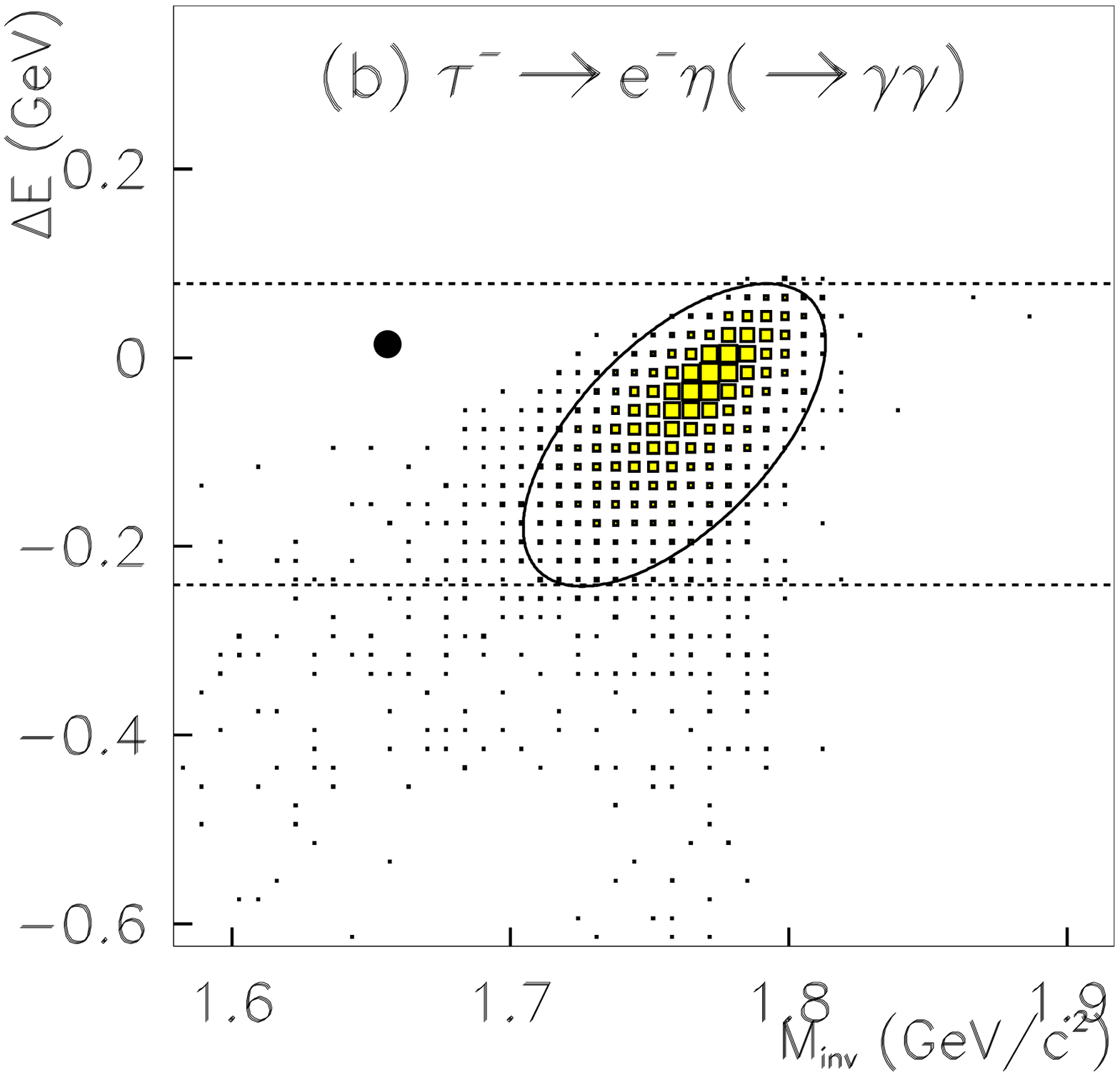}}\\
 \vspace*{-0.5cm} 
 \resizebox{0.3\textwidth}{0.3\textwidth}{\includegraphics
 {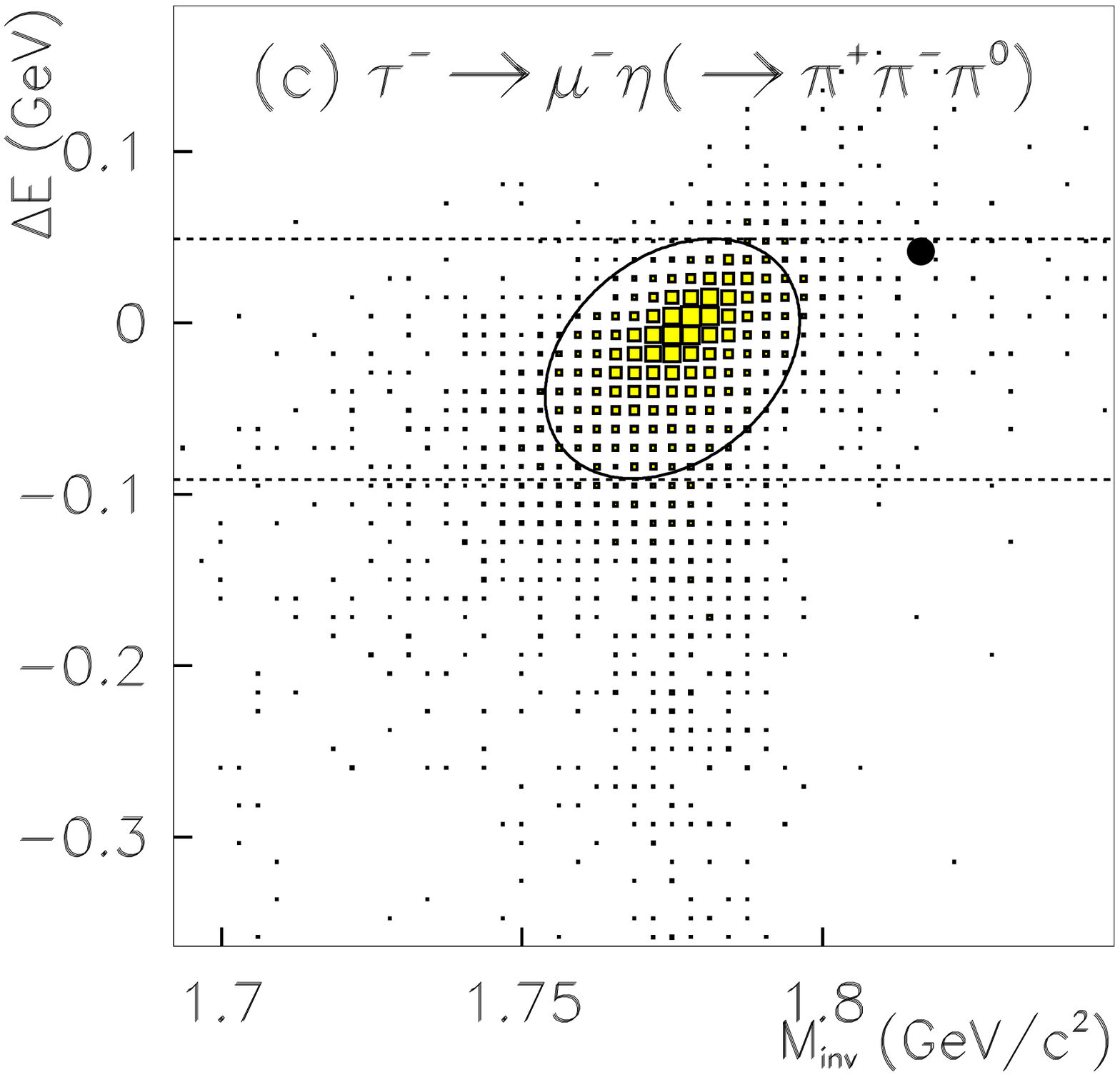}}
 \resizebox{0.3\textwidth}{0.3\textwidth}{\includegraphics
 {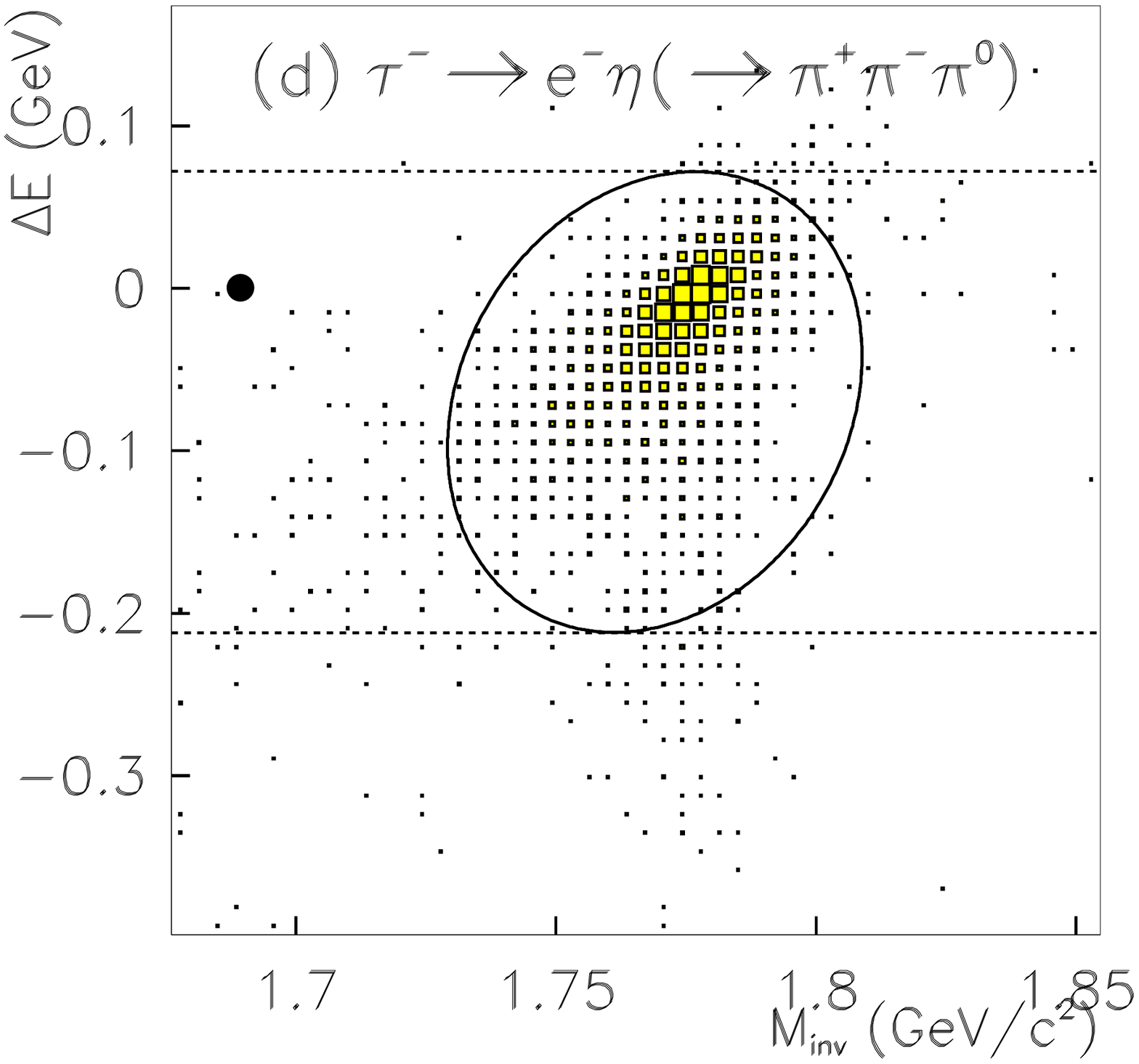}}\\
\vspace*{-0.5cm} 
\caption{
Scatter-plots of data in the
$M_{\rm inv}$ -- $\Delta{E}$ plane:
(a), (b), (c) and (d) correspond to
the $\pm 10 \sigma$ area for
the $\tau^-\rightarrow\mu^-\eta(\to\gamma\gamma)$,
$\tau^-\rightarrow e^-\eta(\to\gamma\gamma)$,
$\tau^-\rightarrow\mu^-\eta(\to\pi^+\pi^-\pi^0)$ and
$\tau^-\rightarrow e^-\eta(\to\pi^+\pi^-\pi^0)$
modes, respectively.
The filled boxes show the MC signal distribution
with arbitrary normalization.
The elliptical signal region shown by the solid curve
is used for evaluating the signal yield.
The region between the horizontal lines excluding the signal region is
used to estimate the expected background in the elliptical region.
}
\label{fig:openbox}
\end{center}
 \end{figure}

\begin{figure}
\begin{center}
 \resizebox{0.3\textwidth}{0.3\textwidth}{\includegraphics
{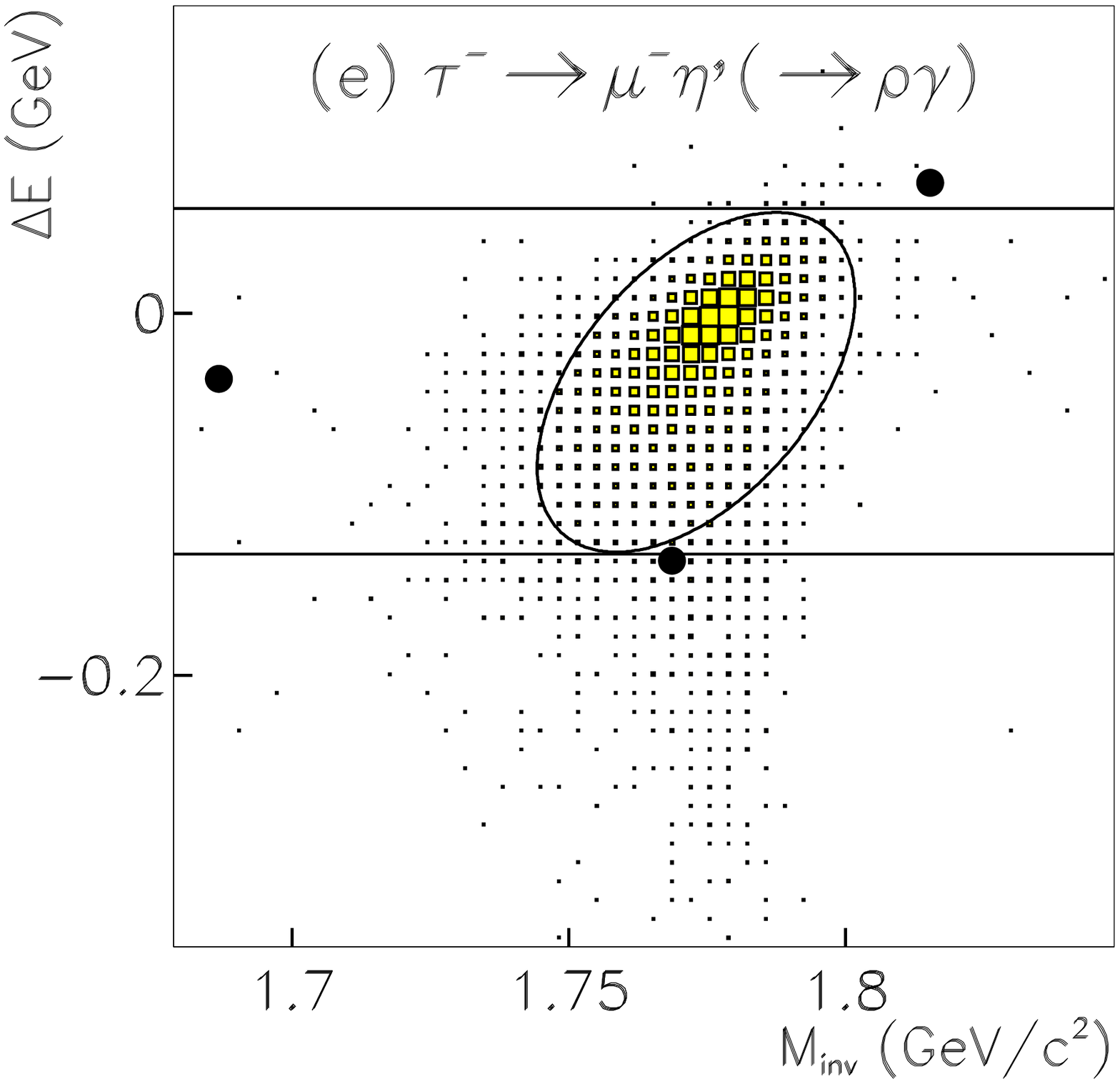}}
 \resizebox{0.3\textwidth}{0.3\textwidth}{\includegraphics
{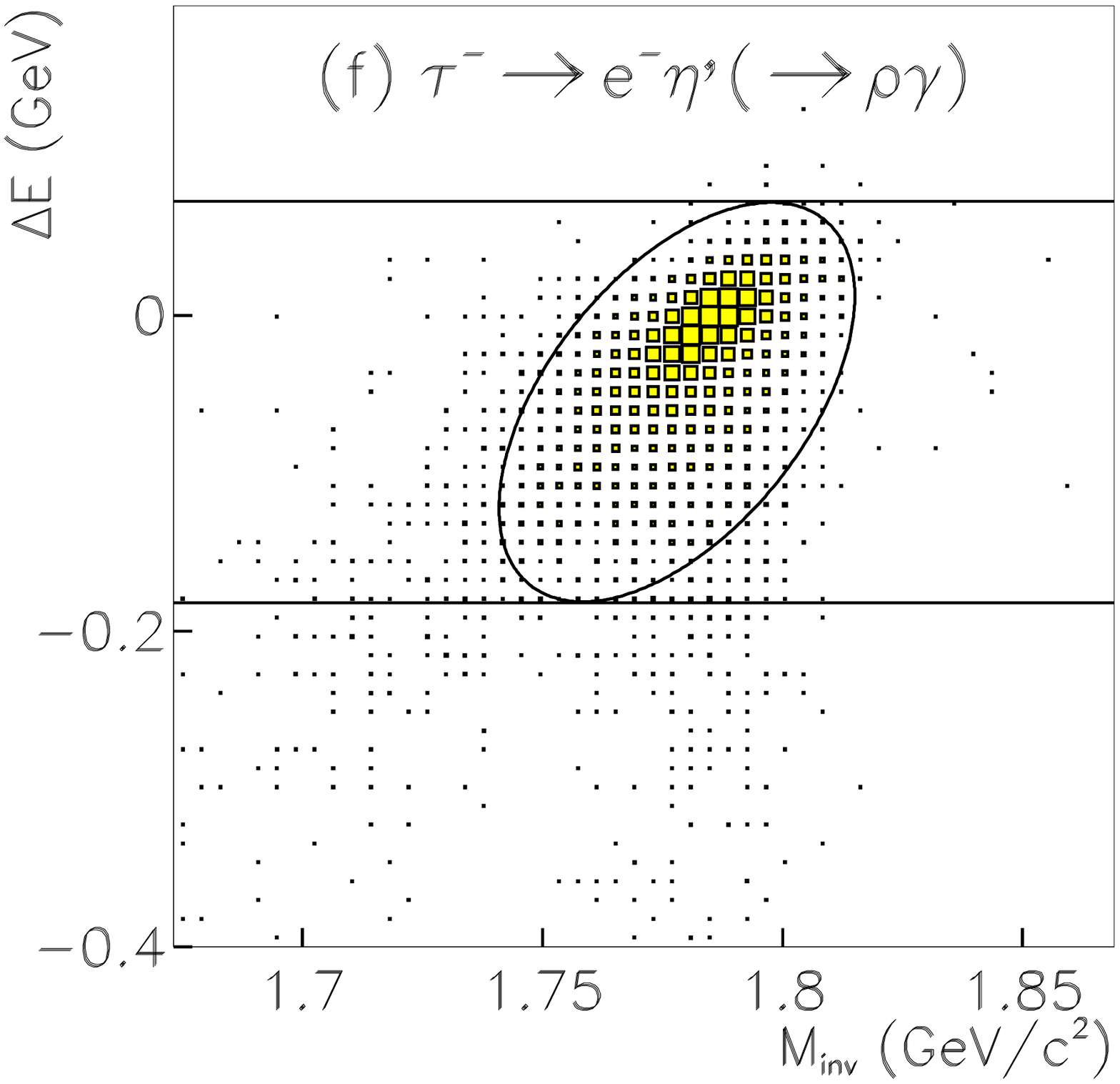}}\\
\vspace*{-0.5cm} 
 \resizebox{0.3\textwidth}{0.3\textwidth}{\includegraphics
{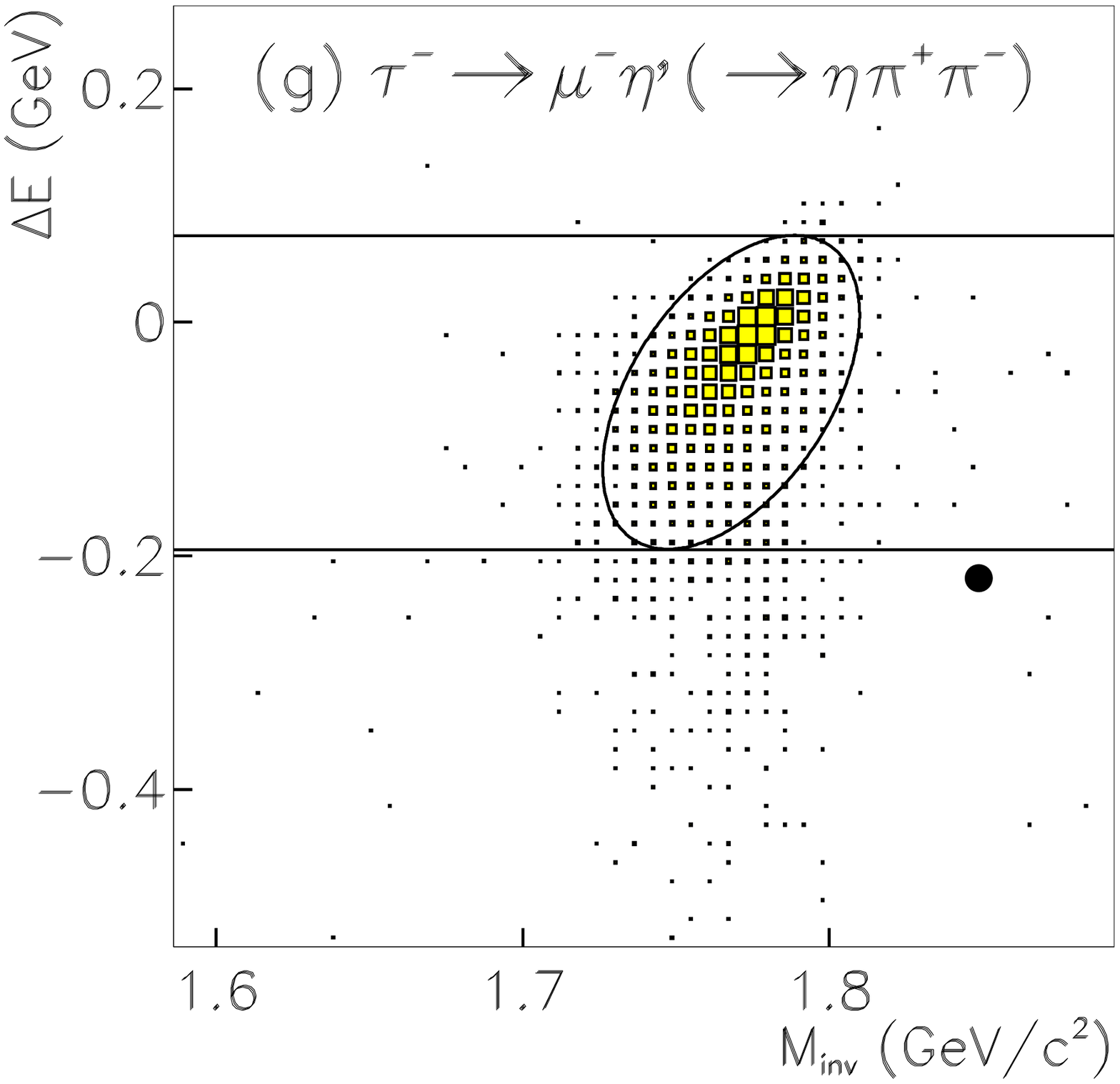}}
 \resizebox{0.3\textwidth}{0.3\textwidth}{\includegraphics
{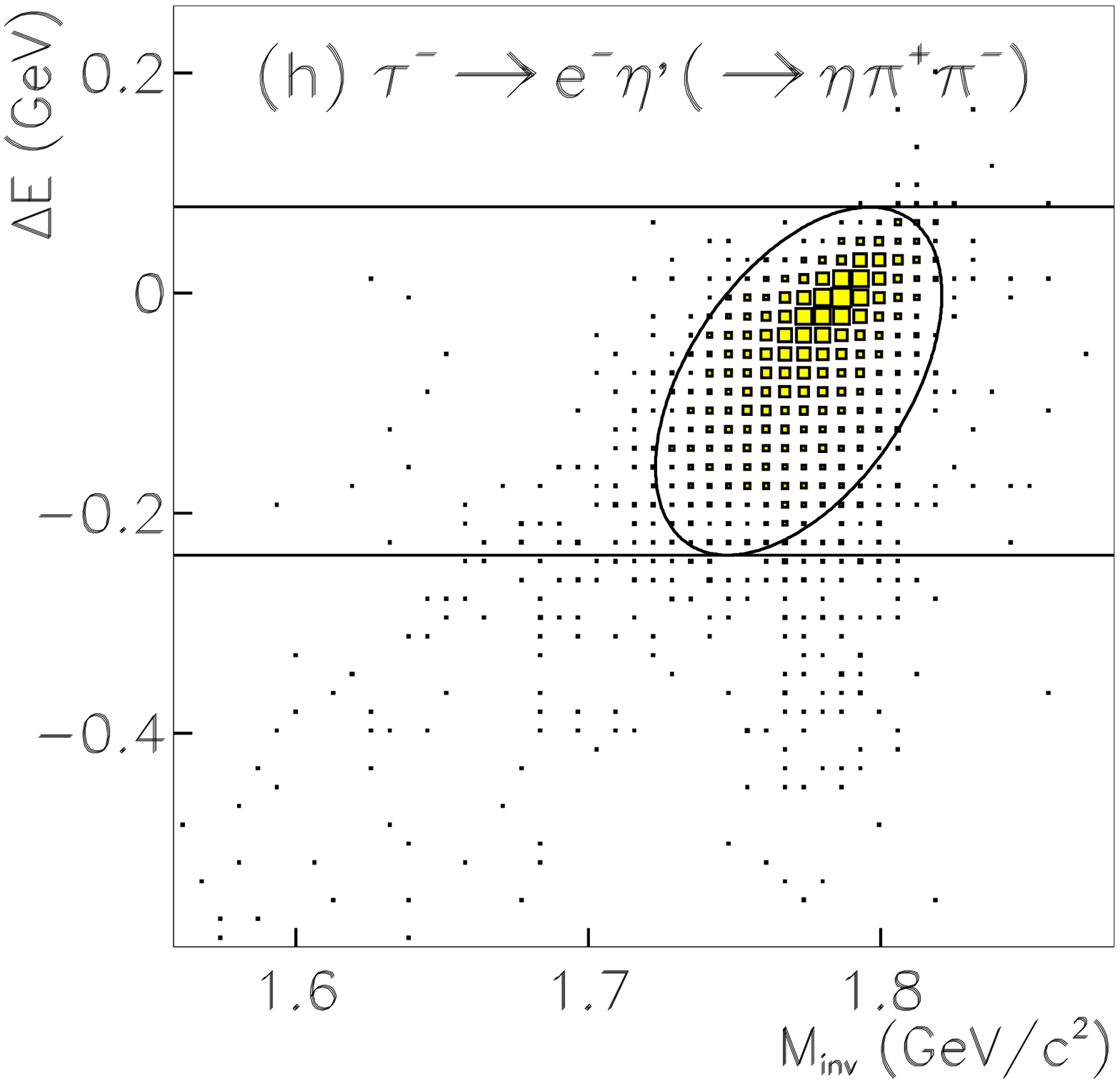}}\\
\vspace*{-0.5cm} 
 \resizebox{0.3\textwidth}{0.3\textwidth}{\includegraphics
{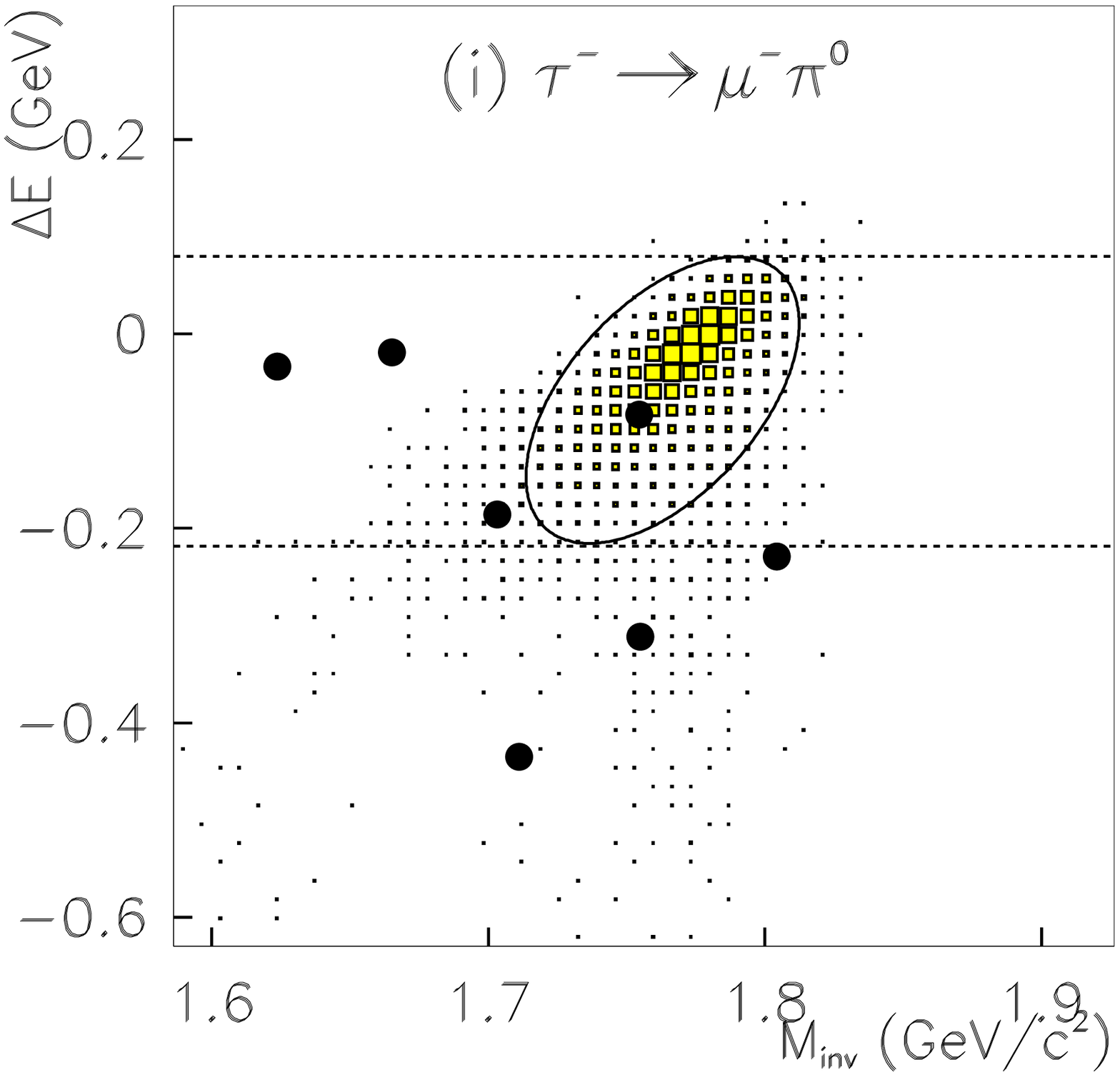}}
 \resizebox{0.3\textwidth}{0.3\textwidth}{\includegraphics
{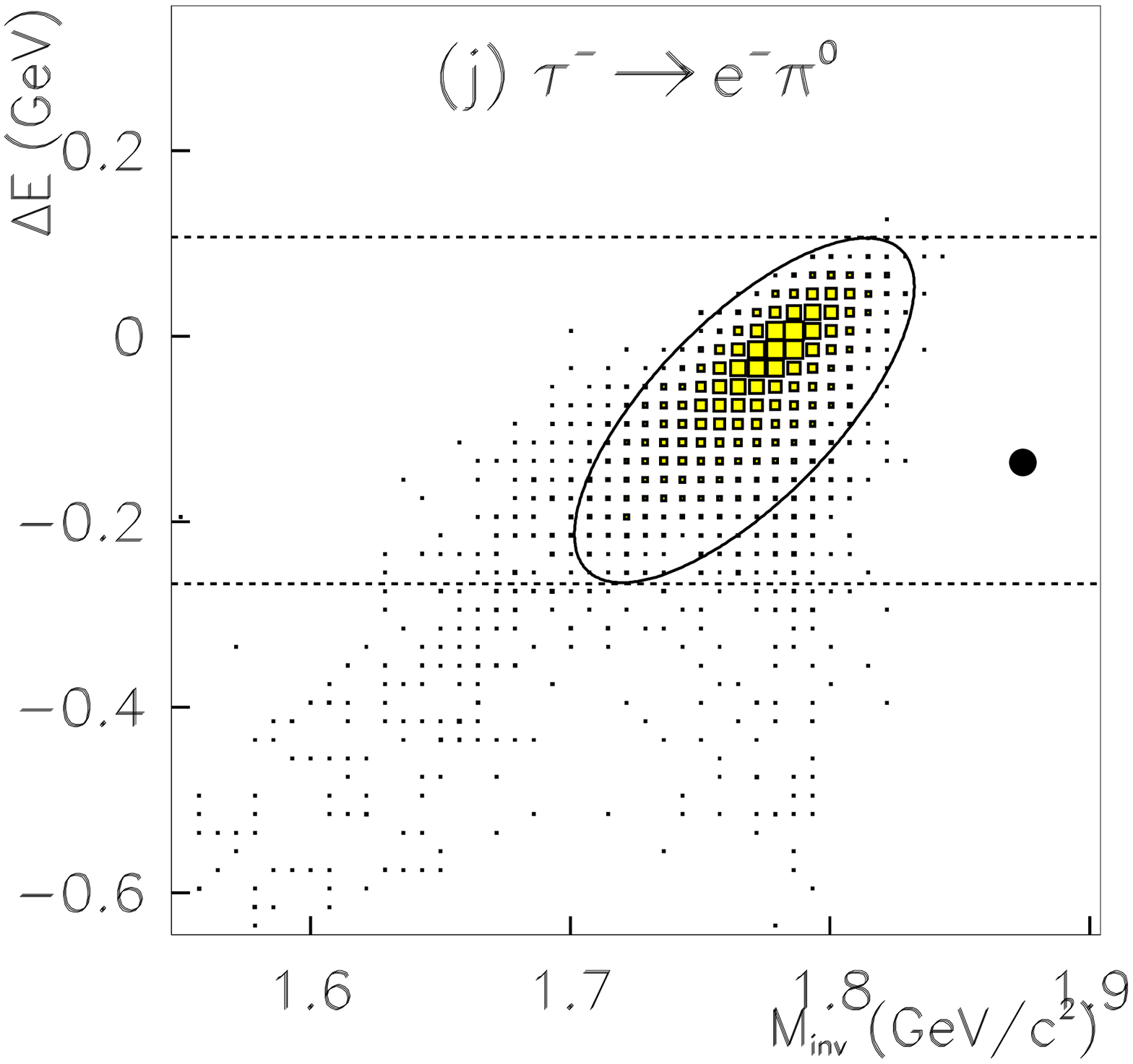}}\\
\caption{
Scatter-plots of data in the
$M_{\rm inv}$ -- $\Delta{E}$ plane:
(e), (f), (g),  (h), (i) and (j) correspond to
the $\pm 10 \sigma$ area for
the $\tau^-\rightarrow\mu^-\eta'(\to\rho\gamma)$,
$\tau^-\rightarrow e^-\eta'(\to\rho\gamma)$,
$\tau^-\rightarrow\mu^-\eta'(\to\eta\pi^+\pi^-)$,
$\tau^-\rightarrow e^-\eta'(\to\eta\pi^+\pi^-)$,
$\tau^-\rightarrow\mu^-\pi^0(\to\gamma\gamma)$ and 
$\tau^-\rightarrow e^-\pi^0(\to\gamma\gamma)$
modes, respectively.
{The data are indicated by the solid circles.}
The filled boxes show the MC signal distribution
with arbitrary normalization.
The elliptical signal region shown by the solid curve
is used for evaluating the signal yield.
The region between the horizontal lines excluding the signal region is
used to estimate the expected background in the elliptical region.
}
\label{fig:openbox2}
\end{center}
\end{figure}

\section{Discussion}

The branching {fraction} {for} the  $\tau^-\to\mu^-\eta$ mode
{may be} 
enhanced by Higgs-mediated LFV 
if large mixing between 
{left-hand
scalar muons and scalar taus} in the corresponding 
SUSY {model 
occurs}~\cite{cite:higgs}.
{This} can be written as
\begin{equation}
{\cal{B}}(\tau^-\to\mu^-\eta) =
8.4\times10^{-7}
\displaystyle\left(
\frac{\tan \beta}{60}
\right)^6
\left(
\displaystyle\frac{100 \mbox{GeV}/c^2}{m_A}
\right)^4,
\end{equation}
where $m_A$ is the pseudoscalar Higgs mass
and 
{$\tan \beta$ is 
the ratio of the vacuum expectation values
of the neutral Higgs fields coupled to 
up-type and down-type fermions.}
From our upper limit {on}
the branching fraction {{for} the decay $\tau^-\to\mu^-\eta$,}
some region of $m_A$ and $\tan \beta$ parameters
can be excluded.
Figure~\ref{fig:mAtanB}
shows the excluded region in the $m_A - \tan \beta$ plane.
{It} also shows
the constraints at {a} 95\% C.L. from the
CDF~\cite{CDF}, $\mbox{D\O}$~\cite{D0} and LEP2 experiments~\cite{LEP}.
{The excluded regions 
from 
{these}
experiments
are
shown
with the Higgs mass parameter 
$\mu > 0$ {in}
the maximum stop-mixing scenario~\cite{Higgs}.}
{We note that 
{their theoretical assumptions 
are
somewhat 
different 
with from ours,}
and thus these regions are for illustrative purposes only.}

{The improved sensitivity to rare $\tau$ lepton decay
{achieved in} this 
{work can also be used} 
to constrain the parameters of 
{other models, e.g.,
{those with}} the heavy Dirac neutrinos~\cite{cite:amon}.
In this model,
the expected branching fractions of various 
{LFV decays} are
evaluated in 
terms of combinations of 
the model parameters.
These parameters, denoted $y_{\tau e}$   and $y_{\tau\mu}$  
for $\tau$ decay involving an electron and a muon, respectively,
can vary from 0 to 1.
We obtain 
the {following} upper limits:
$y_{\tau e} < 0.17$  and $y_{\tau\mu} < 0.47$
at {90\%} C.L. 
from our {$\tau^-\to\ell^-\pi^0$} results.

\begin{figure}
\begin{center}
 \resizebox{0.45\textwidth}{0.45\textwidth}{\includegraphics
{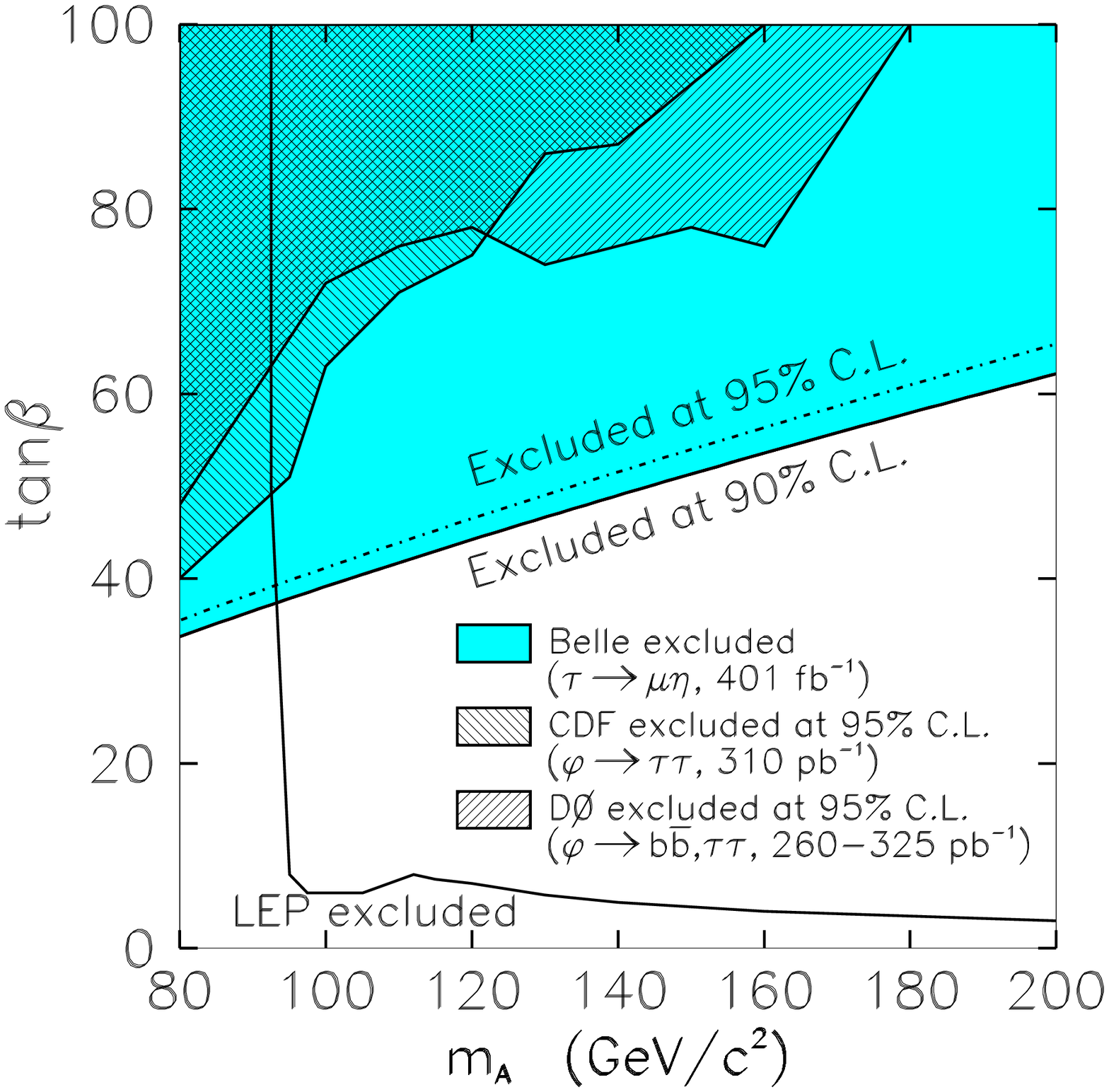}}
 \vspace*{-0.5cm}
\caption{
{The excluded region} 
{in} the $m_A - \tan \beta$ plane
from our results at 90\% C.L.
and
other experiments at 95\% C.L. from
CDF~\cite{CDF}, $\mbox{D\O}$~\cite{D0}, LEP~\cite{LEP}. 
The excluded regions 
from the
CDF, $\mbox{D\O}$ and LEP2 experiments
are
shown
{for $\mu > 0$ in the
maximum stop-mixing scenario~\cite{Higgs}.}
}
\label{fig:mAtanB}
\end{center}
\end{figure}

\section{Summary}

We have searched for {lepton-flavor-violating} $\tau$ 
{decays} with a pseudoscalar meson 
{($\eta$, $\eta'$ and $\pi^0$)}
using 401 fb$^{-1}$ of data.
No signal is found and
we set the following upper limits on branching fractions: 
${\cal{B}}(\tau^-\rightarrow e^-\eta) < 9.2\times 10^{-8}$, 
${\cal{B}}(\tau^-\rightarrow \mu^-\eta) < 6.5\times 10^{-8}$,
${\cal{B}}(\tau^-\rightarrow e^-\eta') < 1.6\times 10^{-7}$, 
${\cal{B}}(\tau^-\rightarrow \mu^-\eta') < 1.3\times 10^{-7}$  
${\cal{B}}(\tau^-\rightarrow e^-\pi^0) < 8.0\times 10^{-8}$
and 
${\cal{B}}(\tau^-\rightarrow \mu^-\pi^0) < 1.2\times 10^{-7}$    
at the 90\% confidence level, respectively. 
{These results improve 
upon 
our previously published upper limits
by factors 
from 2.3 to 6.3.
They are also somewhat better than
the recent results 
from BaBar~\cite{cite:leta_babar}  
with the single exception 
of the limit for the $\tau^-\to\mu^-\pi^0$ mode,
and 
are
the most stringent limits on these modes to 
date.
These limits 
help to constrain new physics scenarios beyond the Standard Model.}

\section*{Acknowledgments}

{We are grateful to A.~Brignole and A.~Rossi for
enlightening discussions.}
We thank the KEKB group for the excellent operation of the
accelerator, the KEK cryogenics group for the efficient
operation of the solenoid, and the KEK computer group and
the National Institute of Informatics for valuable computing
and Super-SINET network support. We acknowledge support from
the Ministry of Education, Culture, Sports, Science, and
Technology of Japan and the Japan Society for the Promotion
of Science; the Australian Research Council and the
Australian Department of Education, Science and Training;
the National Science Foundation of China and the Knowledge
Innovation Program of the Chinese Academy of Sciences under
contract No.~10575109 and IHEP-U-503; the Department of
Science and Technology of India; 
the BK21 program of the Ministry of Education of Korea, 
the CHEP SRC program and Basic Research program 
(grant No.~R01-2005-000-10089-0) of the Korea Science and
Engineering Foundation, and the Pure Basic Research Group 
program of the Korea Research Foundation; 
the Polish State Committee for Scientific Research; 
the Ministry of Education and Science of the Russian
Federation and the Russian Federal Agency for Atomic Energy;
the Slovenian Research Agency;  the Swiss
National Science Foundation; the National Science Council
and the Ministry of Education of Taiwan; and the U.S.\
Department of Energy.


%
%
%

\end{document}